\documentclass[conference]{IEEEtran}
\IEEEoverridecommandlockouts
\usepackage{cite}
\usepackage{amsmath,amssymb,amsfonts}
\usepackage{algorithmic}
\usepackage{graphicx}
\usepackage{textcomp}
\usepackage{xcolor}
\def\BibTeX{{\rm B\kern-.05em{\sc i\kern-.025em b}\kern-.08em
    T\kern-.1667em\lower.7ex\hbox{E}\kern-.125emX}}

\begin{document}

\title{Reinforcement Learning Based Dynamic Function Splitting in Disaggregated Green  Open RANs
{
\thanks{This work is supported by the Turkish Directorate of Strategy and Budget under the TAM Project number 2007K12-873. }
}}

\author{\IEEEauthorblockN{Turgay Pamuklu $\dag$, Melike Erol-Kantarci $\ddag$, \IEEEmembership{Senior Member, IEEE}, Cem Ersoy $\dag$, \IEEEmembership{Senior Member, IEEE} }
\IEEEauthorblockA{\textit{NETLAB, Department of Computer Engineering,}
\textit{Bogazici University}, Istanbul, Turkey \\
\{turgay.pamuklu, ersoy\}@boun.edu.tr}
\IEEEauthorblockA{\textit{School of Electrical Engineering and Computer Science,}
\textit{University of Ottawa}, Ottawa, Canada \\
melike.erolkantarci@uottawa.ca}
}

\maketitle

\makeatletter
\def\ps@IEEEtitlepagestyle{%
  \def\@oddfoot{\mycopyrightnotice}%
  \def\@oddhead{\hbox{}\@IEEEheaderstyle\leftmark\hfil\thepage}\relax
  \def\@evenhead{\@IEEEheaderstyle\thepage\hfil\leftmark\hbox{}}\relax
  \def\@evenfoot{}%
}
\def\mycopyrightnotice{%
  \begin{minipage}{\textwidth}
  \centering \scriptsize
Accepted Paper. 20XX IEEE.  Personal use of this material is permitted.  Permission from IEEE must be obtained for all other uses, in any current or future media, including reprinting/republishing this material for advertising or promotional purposes, creating new collective works, for resale or redistribution to servers or lists, or reuse of any copyrighted component of this work in other works.
  \end{minipage}
}
\makeatother

\begin{abstract}
With the growing momentum around Open RAN (O-RAN) initiatives, performing dynamic Function Splitting (FS) in disaggregated and virtualized Radio Access Networks (vRANs), in an efficient way, is becoming highly important. An equally important efficiency demand is emerging from the energy consumption dimension of the RAN hardware and software. Supplying the RAN with Renewable Energy Sources (RESs) promises to boost the energy-efficiency. Yet, FS in such a dynamic setting, calls for intelligent mechanisms that can adapt to the varying conditions of the RES supply and the traffic load on the mobile network. In this paper, we propose a reinforcement learning (RL)-based dynamic function splitting (RLDFS) technique that decides on the function splits in an O-RAN to make the best use of RES supply and minimize operator costs. We also formulate an operational expenditure minimization problem. We evaluate the performance of the proposed approach on a real data set of solar irradiation and traffic rate variations. Our results show that the proposed RLDFS method makes effective use of RES and reduces the cost of an MNO. We also investigate the impact of the size of solar panels and batteries which may guide MNOs to decide on proper RES and battery sizing for their networks.
\end{abstract}


\section{Introduction}
\begin{figure*}[t]
\centering
\includegraphics[width=0.7\textwidth]{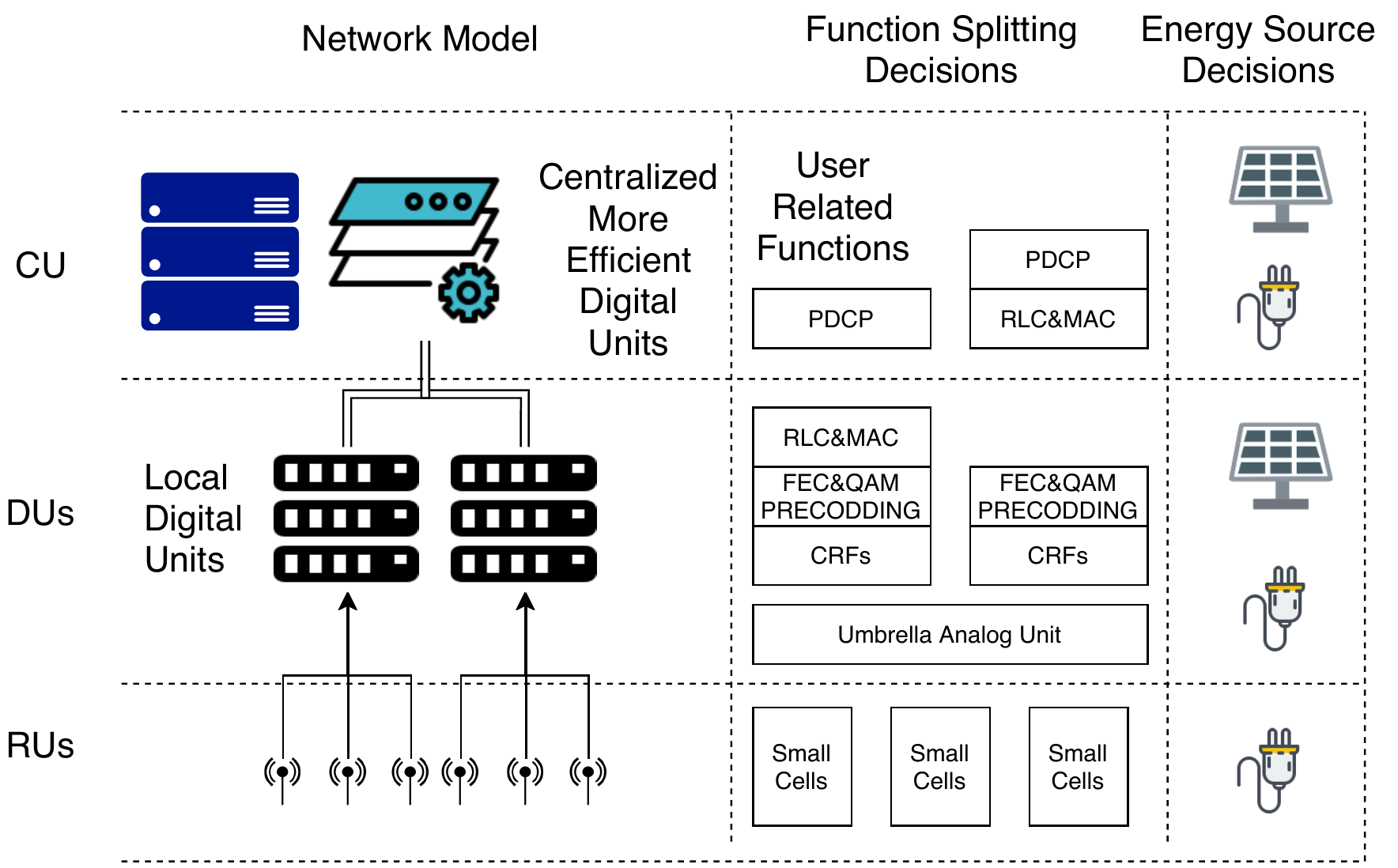}
\caption{\label{fig:arch} Dynamic function splitting between energy harvesting telco clouds. A Mobile network operator has to deal with two cost-related decisions  (CU: central unit, DU: distributed unit, RU: radio unit, PDCP: packet data convergence protocol, RLC: radio link control, MAC: medium access control, FEC: forward error correction, QAM: quadrature amplitude modulation, CRF: cell-related functions).}
\end{figure*}

\par The ideas of virtualized RANs (vRANs) and functional splits date back to Small Cell Forum's studies on small cell virtualization \cite{SCF2015}. Accordingly, vRAN mainly aims to disaggregate Baseband Units (BBU)  and remote radio head (RRU) functionalities and allow softwarized network functions of BBUs to be hosted on common-of-the-shelf (COST) hardware. The new Open Radio Access Network (O-RAN) architecture aims to define open interfaces in disaggregated vRANs such that equipment from multiple vendors can be interoperable \cite{O-RANAlliance2020}. Within the O-RAN architecture, horizontal and vertical splits allow network functions to be handled by either the hardware at the edge cloud, known as Distributed Unit (DU), or the hardware at the central cloud, i.e. the centralized unit (CU). 
\par In O-RAN, function splitting (FS) can be either between physical and virtualized resources, or between DU and CU, yielding many possibilities. Yet, the full potential of FS arises from dynamic splits where network functions are placed based on varying feedback from the network. The traditional user Quality of Service (QoS) is certainly central to decision making in dynamic FS. For instance, the traffic type, whether it is enhanced mobile broadband (eMBB) or ultra-reliable low-latency communication (URLLC), impacts the FS decision. In addition, energy efficiency plays a key role in a mobile network operator's (MNO) decision-making process due to the tremendous growth in their carbon footprint and cost in relation to densification and data demand \cite{Pamuklu2020}. Using renewable energy sources (RESs) in telco clouds as an alternative to on-grid energy is a promising approach to reduce the excessive energy consumption from the grid. On the other hand, renewable energy has two critical drawbacks. First, MNOs should store this energy in a battery that has limited capacity. Due to the increasing CapEx of RES and battery, MNOs need to optimize energy usage \cite{Pamuklu2018}. Second, RES is intermittent and the supply has some level of uncertainty \cite{Chiang2018}. Therefore, under the variability of the RES and the variability of the wireless environment, optimal functional splitting in disaggregated green virtualized RANs introduces a great degree of complexity which can be best addressed by machine learning techniques, in particular reinforcement learning-based methods \cite{Temesgene2019}. 
\par This paper focuses on a novel system model in which the functions are split between a CU and several DUs while using the RES as an alternative energy source for the telco cloud. Figure~\ref{fig:arch} presents this architecture. We propose a reinforcement learning-based (RL) dynamic function splitting (RLDFS) approach and implement Q-learning and SARSA algorithms. Our motivation to choose these algorithms originates from their light-weight implementation and their dynamically learning properties. Thus, our studied network can conveniently adapt to the continuous traffic and solar radiation changes. We experiment with various densities of traffic loads and real solar energy data collected at different seasons to see the impact of seasonal changes in four globally separate geographical areas, which have significantly diverse solar radiation distributions. Our results show that the proposed solution reduces the OpEx of an MNO significantly under various solar radiation and traffic load.

\par The remainder of this paper is organized as follows. Section 2 summarizes the related work. We define the system model and the cost optimization problem in Section 3. In Section 4, we explain the RLDFS technique and present its performance in Section 5. Section 6 concludes the paper.
\section{Related Work}
\par Temesgene et al. have proposed the first studies that merge the FS approach and RESs usage in a RAN. The authors detail the energy consumption in virtual small cells (vSCs) and implement reinforcement learning (RL) methods to optimize the FS decisions between the vSCs and a macro base station \cite{Temesgene2018,Temesgene2019}. Meanwhile, Wang et al. propose a heuristic solution for an architecture that the functions that are split between the RRHs and BBUs \cite{Wang2018}. In addition, Ko et al. focus on a similar architecture and formulate a constrained Markov decision process model \cite{Ko2018}. Nevertheless, all of these studies aim to split the functions between the RRHs and a BBU center.
\par Larsen et al. provide a broad survey for both static and dynamic FS methods. Also, it is among the first research works that consider O-RAN architecture \cite{Larsen2019}. Furthermore, Alabbasi et al. introduce a study called Hybrid Cloud RAN where they aim to jointly minimize the energy and bandwidth consumption in their RAN with a constrained programming solver  \cite{Alabbasi2019}. Their results highlight the importance of FS decisions for delay and energy consumption minimization.  However, their study does not consider using RESs usage in the RAN. 
\par In our previous study, we provide mixed-integer programming solver and heuristic solutions for this joint problem \cite{Pamuklu2020}. Although the new study is based on this paper and aimed to minimize the OpEx, the previous research considers a hybrid C-RAN architecture and includes a bin-packing problem due to function-micro data center assignment decisions. Different than prior work, in this paper, we decide the function splittings based on the user/traffic types (URLLC/eMBB) tuples, and we focus on a novel O-RAN architecture. Also, we choose two RL-techniques, Q-Learning and SARSA, as solution techniques. 

\section{System Model}
\par We consider a virtualized RAN environment where network functions can be split dynamically between CU and DU, based on the network conditions. This allows for more flexibility than the fixed functional split options provided by 3GPP. Figure~\ref{fig:arch} illustrates a three-tier vRAN model that employs radio unit (RU), DU and CU, in addition to being green by having solar power generation capability at DUs and CUs. 
\par We may classify the network functions into two groups: The user-related functions (URFs) dedicated to specific user data and cell-related functions (CRFs) that process the multiplexed URFs in the physical layer \cite{Wang2017}. The top-tier, CU, has energy efficient digital units to execute only URFs. This CU has directed fiber optic links to the middle-level DUs ($r\in\mathcal{R}$). DUs are responsible for both URFs and CRFs. There are two reasons to prevent handling the CRFs at the CU level. First, processing the CRFs at this level need huge bandwidth allocation at the fiber optic links \cite{SmallCellForum2016}. Second, operating these functions at the DUs provides relaxation for the stringent delay requirements \cite{Pamuklu2020}. 
Finally, a set of RUs are connected to their corresponding DU with a point-to-point millimeter-wave or dedicated fiber link \cite{Alabbasi2019}. 
In this architecture, the MNO can deploy the RUs near to their corresponding DU to hold their capital expenditure (CapEx) at more economical rates. Furthermore, as DUs are located closer to the users, they provide latency advantage over CUs for certain types of traffic.

\par Deciding the optimum splits for network functions ($f\in\mathcal{F}$) is the fundamental goal in this system.  The split decisions should be dynamically made for a set of time intervals in a day ($t\in\mathcal{T}$) for each traffic type ($i\in\mathcal{I}$), such as eMBB or URLCC. The other significant decision problem is related to using two different energy sources at the DU and CU. The first energy source, a solar panel, reduces the OpEx of the MNO with renewable energy; the second one, on-grid energy, acts as reliable energy source in the case of insufficient green energy. Before formulating the relations between these two types of energy sources, we will describe the total energy consumption in the DU ($E^{DU}_{rt}$) and the CU ($E^{CU}_{t}$) determined by eqs. (~\ref{eq:energyConsInDU}) and (\ref{eq:energyConsInCU}), respectively:
\begin{align}
E^{DU}_{rt} &= \left[E^{DU}_{S} + \sum\limits_{i\in\mathcal{I}} \sum\limits_{f\in\mathcal{F}} U_{rit} * a_{ritf} * E^{DU}_{D} \right]
\label{eq:energyConsInDU}
\\
E^{CU}_{t} &= \left[E^{CU}_{S} + \sum\limits_{r\in\mathcal{R}}\sum\limits_{i\in\mathcal{I}} \sum\limits_{f\in\mathcal{F}} U_{rit} * (1 - a_{ritf}) * E^{CU}_{D}\right]
\label{eq:energyConsInCU}
\end{align}  
\par $E^{DU}_{S}$ is the static energy consumption of a DU due to the cooling system and the other idle-mode energy consumption which do not change by the function split decisions. A CU also has a static energy consumption ($E^{CU}_{S}$) due to its idle-mode energy consumption. Besides, dynamic energy consumption has three main components. The first component $U_{rit}$ is the traffic load of traffic type $i$ at the DU $r$ in time slot $t$. The second component, $a_{ritf}$, is a binary decision variable that equals to one if the URF $f$ of traffic type $i$ is executed at the DU $r$ in time slot $t$. The last one is the dynamic energy coefficient, represented as $E^{DU}_{D}$ and $E^{CU}_{D}$ for the DU and CU, respectively. Lastly, the energy consumption of CU (eq. (\ref{eq:energyConsInCU})) depends on the number of functions that are not processed in DU $(1-a_{ritf})$ and the traffic loads of these functions ($U_{rit}$). Thus, it is crucial to optimize the FS decisions to reduce overall energy consumption.
\par The relation between the two types of energy sources can be formulated with the following OpEx minimization problem as in eq. (\ref{j3:eq:obj1}).
The OpEx of the system is the overall on-grid electricity bills of the CU and the DUs. Note that the energy consumption of the RU and the cost of solar generation, in terms of investment, are not included in OpEx calculation. The reason is that these costs do not impact the dynamic FS problem. 

In this model, we consider three fundamental problems to reduce the energy bill of an MNO. First, we have to reduce the total energy consumption with the efficient splitting of the URFs. Second, we have to promote the usage of solar energy ($p^{CU}_{t}$ and $p^{DU}_{rt}$) instead of on-grid energy. Hence, the amount of surplus solar energy needs to be minimized, and in relation, the capacity for solar generation and batteries needs to be selected such that the cost of MNO is minimized. Lastly, we have to consider the variation of the electricity prices in a day period ($c^{E}_{t}$) and use renewable energy during peak electricity rates.

\begin{align}
&\textbf{Minimize:} \notag \\
&\mathbb{O} = \sum\limits_{t\in\mathcal{T}}  \biggl [  E^{CU}_{t}-p^{CU}_{t} +  \sum\limits_{r\in\mathcal{R}}(E^{DU}_{rt}-p^{DU}_{rt}) \biggr] * c^{E}_{t}
\label{j3:eq:obj1}
\\
&\textbf{Subject to:}\notag \\
&(1 - a_{ritf_{x}}) * \sum\limits_{f_{y} \ge f_{x}}^{f_{y}\in\mathcal{F}} a_{ritf_{y}} = 0,\forall f_{x} \in \mathcal{F}, \forall i \in \mathcal{I}, \forall r \in \mathcal{R}
\label{eq:FunctionAssign}
\\
&b^{CU}_{t} = b^{CU}_{(t-1)} - p^{CU}_{t} + \omega^{CU}G^{CU}_{t}
\label{eq:battEnergyInCU}
\\
&b^{DU}_{rt} = b^{DU}_{r(t-1)} - p^{DU}_{rt} + \omega^{DU}_{r}G^{DU}_{rt},\quad\forall r\in\mathcal{R}
\label{eq:battEnergyInDU}
\\
&b^{CU}_{t} \leq \beta^{CU}
\label{eq:battLimitInCU}
\\
&b^{DU}_{rt} \leq \beta^{DU}_{r},\quad\forall r\in\mathcal{R}
\label{eq:battLimitInDU}
\\
&p^{CU}_{t} \leq  E^{CU}_{t}
\label{eq:renEnMaxLimitInCU}
\\
&p^{DU}_{rt} \leq  E^{DU}_{rt},\quad\forall r\in\mathcal{R}
\label{eq:renEnMaxLimitInDU}
\end{align}

\par Here, eq. (\ref{eq:FunctionAssign}) guarantees that the URFs chain is broken only once between the DUs and the CU. Note that each constraint should be satisfied for all time intervals ($\forall t \in \mathcal{T}$). Assume that the function $f_{x}$ is executed at the CU, then $a_{ritf_{x}}$ equals to zero. After that, the remaining upper layer functions $f_{y}$ should also be completed at the CU ($a_{ritf_{y}}=0$); otherwise, the multiplication on the left side of the equation will be different from zero.
\par Equations~\ref{eq:battEnergyInCU} to~\ref{eq:renEnMaxLimitInDU} regulate the energy usage by the servers in CU and DU. The first two equations determine the renewable energy in the batteries of the solar generators that sit either beside DU or CU. For the CU side (eq.(\ref{eq:battEnergyInCU})), the current battery energy ($b^{CU}_{t}$) depends on the battery's remaining energy from the previous time interval ($b^{CU}_{(t-1)}$). This value increases with the size of the solar panel at the CU ($\omega^{CU}$), and the generated renewable energy of a panel's unit size ($G^{CU}_{t}$). On the other hand, the energy in the CU battery reduces with the consumed green energy ($p^{CU}_{t}$). The calculation is similar for the DU side, which is provided by eq. (\ref{eq:battEnergyInDU}). Meanwhile, the batteries' capacities ($\beta^{CU}$ and $\beta^{DU}_{r}$) limit the maximum stored renewable energy presented by eqs. (\ref{eq:battLimitInCU}) and (\ref{eq:battLimitInDU}) for the CU and DUs, respectively. Lastly, it is clear that the consumed green energy ($p^{CU}_{t}$ and $p^{DU}_{t}$) should be lower than the total energy consumption in a specific time interval, and this constraint is guaranteed by eqs. (\ref{eq:renEnMaxLimitInCU}) and (\ref{eq:renEnMaxLimitInDU}). Solving this optimization model for each arriving user traffic, and under varying solar conditions is not practical. Instead, we propose to use reinforcement learning to allow the network to adapt to dynamic conditions. 
\begin{figure}
\centering
\includegraphics[width=0.40\textwidth]{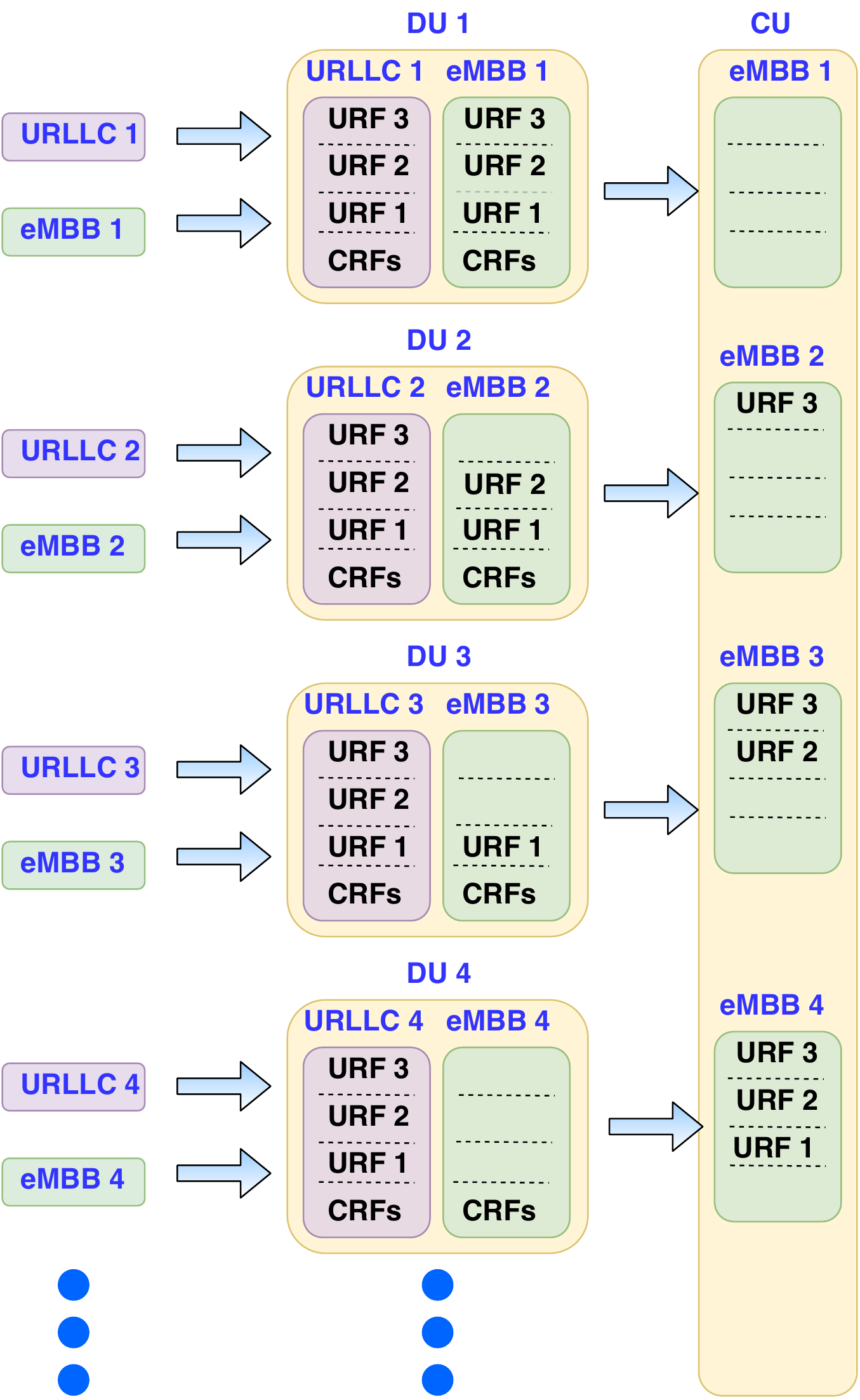}
\caption{\label{fig:top} Traffic type (eMBB or uRLLC) based function splitting in the virtualized open RAN.} 
\end{figure}  

\section{Reinforcement Learning Based Dynamic Function Splitting (RLDFS)}
\par In our model, we use a multi-agent system where DUs and CU act as independent agents. The states for a DU are calculated with the tuple, $S^{DU}_{t}=\{b^{DU}_{rt}, U_{rit}, t^{D}\}$, which includes the remaining energy in the DU's battery, the traffic load on this DU and the time of the day ($t^{D}=t\pmod{T^{D}}$), respectively. Meanwhile, CU states are calculated with the tuple, $S^{CU}_{t}=\{b^{CU}_{rt}, \sum\limits_{r\in R} U_{rit}, t^{D}\}$, which includes the remaining energy in the CU's battery, the total traffic load on the network, and the time of the day, respectively. There are two main benefits to include the time of the day as a state parameter. First, we want to combine the impact of solar irradiance and the electricity price tariffs together. Second, we want to reduce the state space to improve the convergence. Otherwise, we would have to specify individual states for different solar data and electricity prices. Combining this approach with the multi-agent system, we limit the state space to a reasonable size, which yields fast converging rates for the optimal functional split decisions that need to be determined in hours timescale.

\par The OpEx minimization problem has two important decision variables. The actions of the DU agents render them with the tuple $A^{DU}_{t}=\{a_{ritf}, p^{DU}_{rt}\}$. Meanwhile, the CU agent needs to decide only its renewable energy consumption ($A^{CU}_{t}=\{p^{CU}_{t}\}$). In a real-world scenario, the CU automatically processes the URFs not chosen to be processed at the DUs. Finally, the reward is calculated by eq. (~\ref{eq:reward}), in which $\mathbb{O}_{t^{'}}$ is the OpEx in time slot $t^{'}$ , $\psi$ is the normalization factor, and $\tau$ is the window size. 

\begin{flalign}
R_{t+1} &= - \psi \sum\limits_{t^{'}=t-\tau}^{t}\mathbb{O}_{t^{'}}&
\label{eq:reward} \\
Q(S_{t},A_{t}) &\leftarrow Q(S_{t},A_{t}) + \alpha \biggl [ R_{t+1}& \notag \\ 
&+\gamma \max\limits_{A} Q(S_{t+1},A) - Q(S_{t},A_{t})  \biggr]&
\label{eq:qlearning}\\
Q(S_{t},A_{t}) &\leftarrow Q(S_{t},A_{t}) + \alpha \biggl [ R_{t+1}& \notag \\ 
&+\gamma Q(S_{t+1},A_{t+1}) - Q(S_{t},A_{t})  \biggr]&
\label{eq:sarsa}
\end{flalign}
\par We solve the OpEx minimization problem with two RL approaches. The first one, called Q Learning, is represented by eq. (\ref{eq:qlearning}). The second one, called Sarsa, is represented by eq. (\ref{eq:sarsa}). The main difference between these two methods relies on the calculation of the next q-table ($S_{t},A_{t}$). The first one promotes the action that provides the maximum q-value; the second one applies the next action ($A_{t+1}$) directly into account to calculate the next q-table \cite{Sutton2018}. Otherwise, the essential detail to implement an RL solution is to decide the states $S_{t}$, the actions $A_{t}$ and the reward calculation $R_{t+1}$.

\section{Performance Evaluation}
\begin{figure}
\centering
\includegraphics[width=0.45\textwidth]{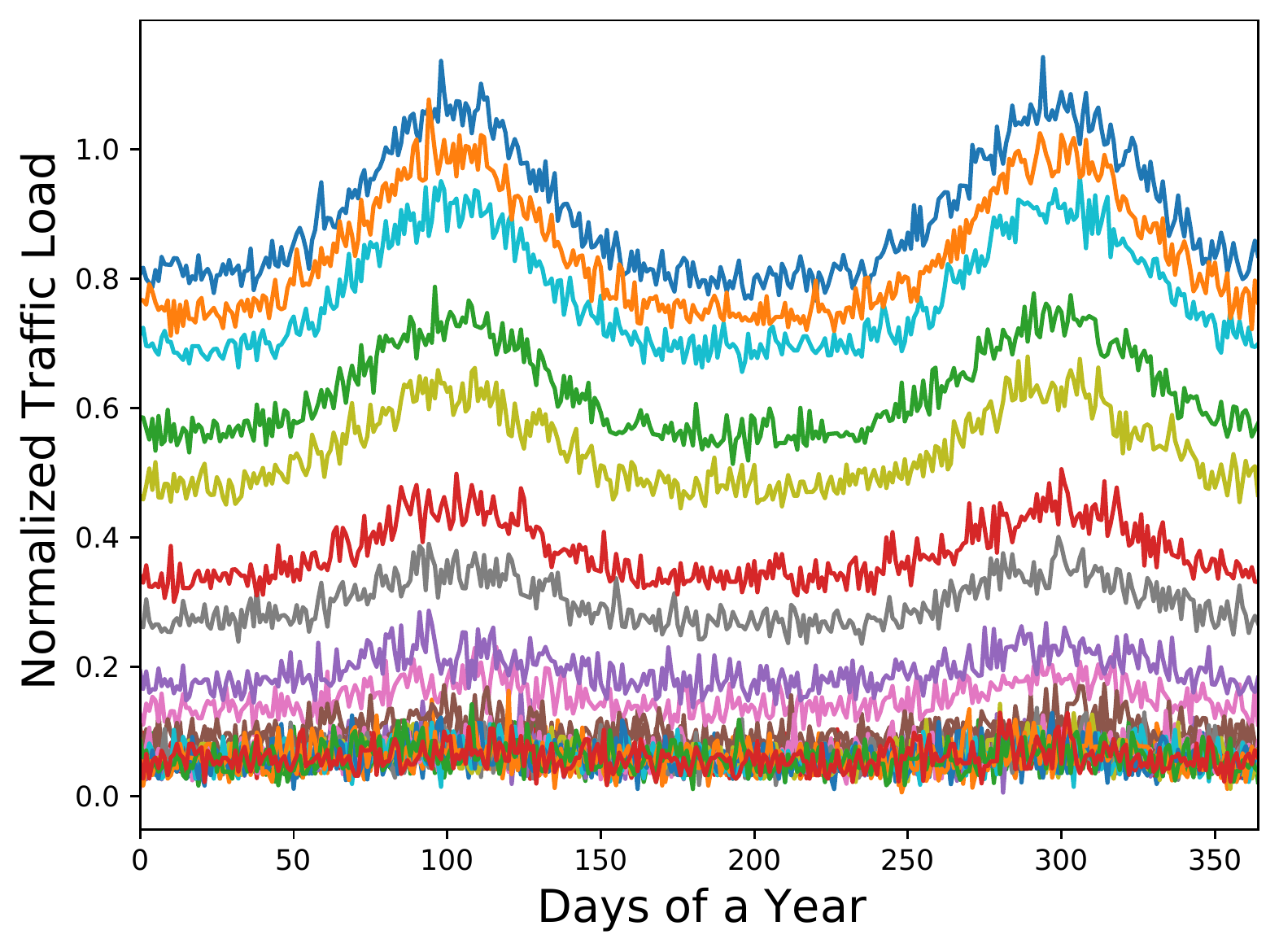}
\caption{\label{fig:tload} eMBB traffic load in a year period. Each legend represents an hour of a day.}
\end{figure}  
\begin{table}[b]
\centering
\caption{\label{tab:Parameters} Energy Parameters.}
\begin{tabular}{c|c|c|c}
Instance & Unit & CU Side & DU Side \\ \hline
$E_{S}$ & kWh & 10 &5 \\
$E_{D}$ & kWh & 0.9 & 1 \\
$\omega$ & kWh & 500 &100 \\
$\beta$ & kWh & 500 &100 \\
$c^{E}$ & \$ & [0.03, 0.07, 0.11] &[0.03, 0.07, 0.11] \\
\end{tabular}
\end{table}
\begin{table}
\centering
\caption{\label{tab:rlparam} RL Parameters.}
\begin{tabular}{c|c|c }
Explanation &  Not. & Value \\ \hline
Number of Episodes & $N^{EP}$ & 4000 \\
Learning Rate & $\alpha$ & 0.05 \\
Discount & $\gamma $ & 0.90 \\
Exploration Starting Value& $\epsilon$ & 0.5 \\
Exploration Decay Value& $\epsilon^{D}$ & 5E-5 \\
Reward Window Size& $\tau$ & 48 \\
\end{tabular}
\end{table}
\begin{figure*}
\centering
\includegraphics[width=0.59\textwidth]{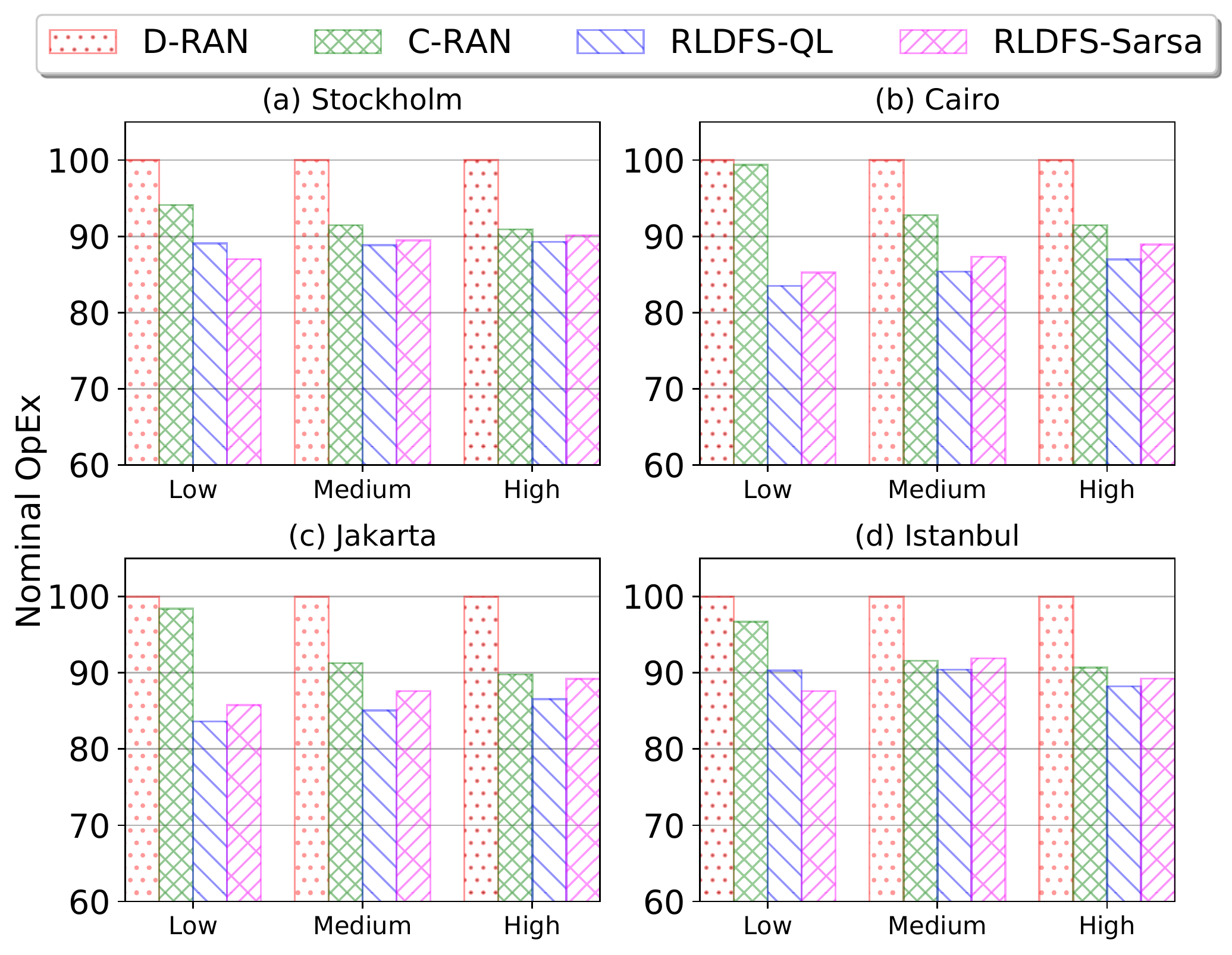}
\caption{\label{fig:tscity} Comparison of RLDFS-QL and RLDFS-Sarsa methods with centralized RAN and distributed RAN approaches for different cities and traffic rates.}
\end{figure*}
\par The evaluation setting of our primary use case is shown in Figure~\ref{fig:top}. We have 1 CU and 20 DUs that each one has one solar panel and a battery to store harvested renewable energy. There are 10000 users serviced by each DU. These users generate two types of traffic, i.e. URLLC and eMBB. The URLLC traffic demands low latency; thus, they are processed directly on DUs \cite{Elsayed2019}. On the other hand, the functions of eMBB traffic, which creates a ten times larger load than the URLLC, are split between the DUs and the CU. 
\begin{equation}
\label{eq:trafficCreator}
\lambda_{it} = \frac{1}{2^{\nu}}[1+\sin(\pi t/12 + \varphi )]^{\nu} + n(t), i\in\mathcal{I}
\end{equation}
\par The users have a daily sinusoidal shape traffic load described by eq. (\ref{eq:trafficCreator}) in which $\varphi$ is a random value between the $3\pi/4$ and $7\pi/4$ which defines the peak hour of the traffic profile, $\nu=7$ determines the slope of the traffic profile and $n(t)$ is a random value which produces a fluctuation in this traffic profile \cite{Zhang2013}. We also provide a yearly load variation to understand the impact of the distinction between the seasons (Figure~\ref{fig:tload}). In addition, we generate different peak hours for each DU to affect distinctive zones in a city such as residential, industrial, or shopping areas \cite{Xu2017,Pamuklu2018}. Thus, we simulate both temporal and spatial variations of a traffic load in the region of a city.

\begin{figure}
\centering
\includegraphics[width=0.50\textwidth]{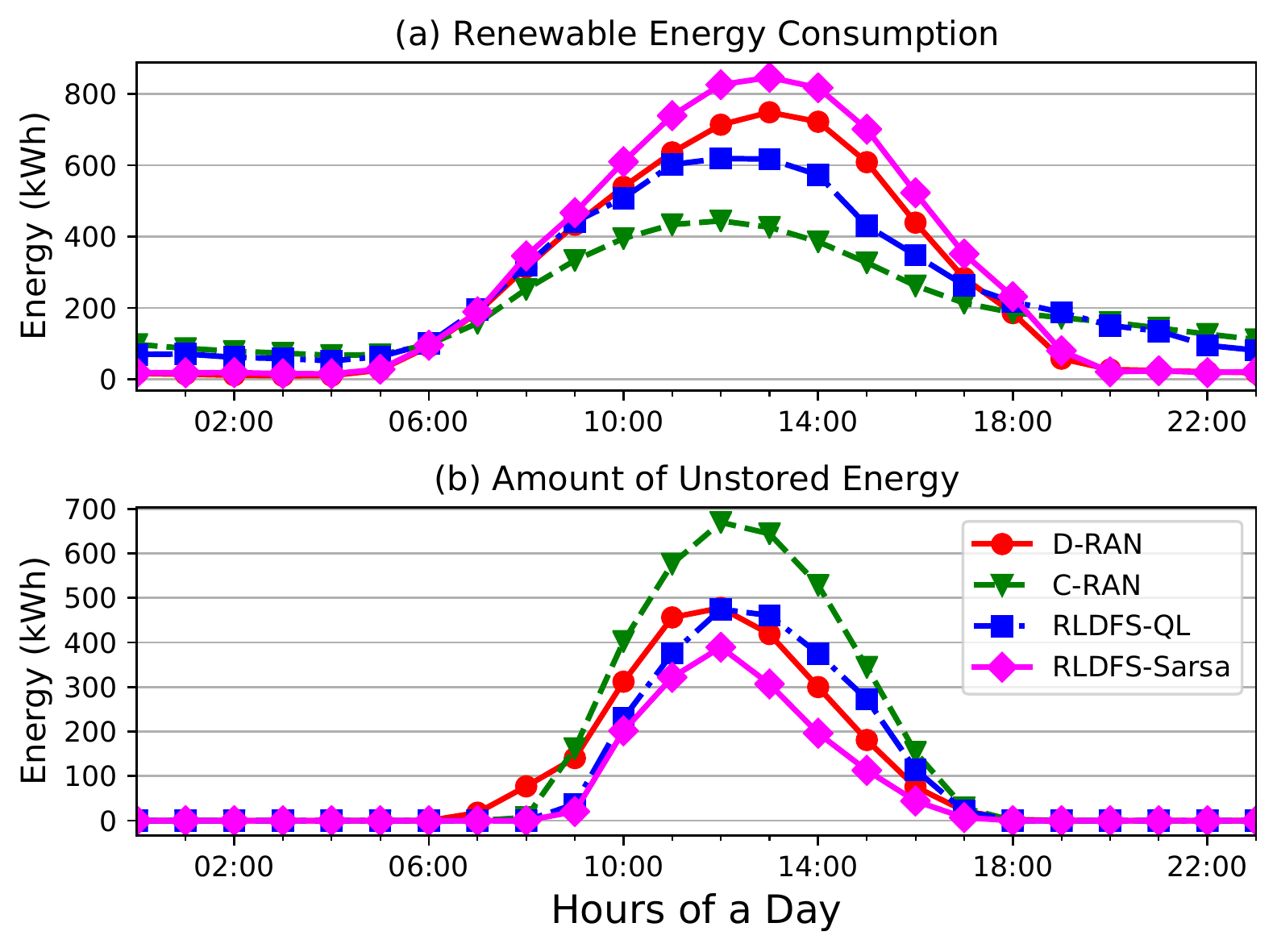}
\caption{\label{fig:daily} Daily distribution of renewable energy consumption and unstored energy in Jakarta-Medium Traffic.}
\end{figure}  

\begin{figure}
\centering
\includegraphics[width=0.50\textwidth]{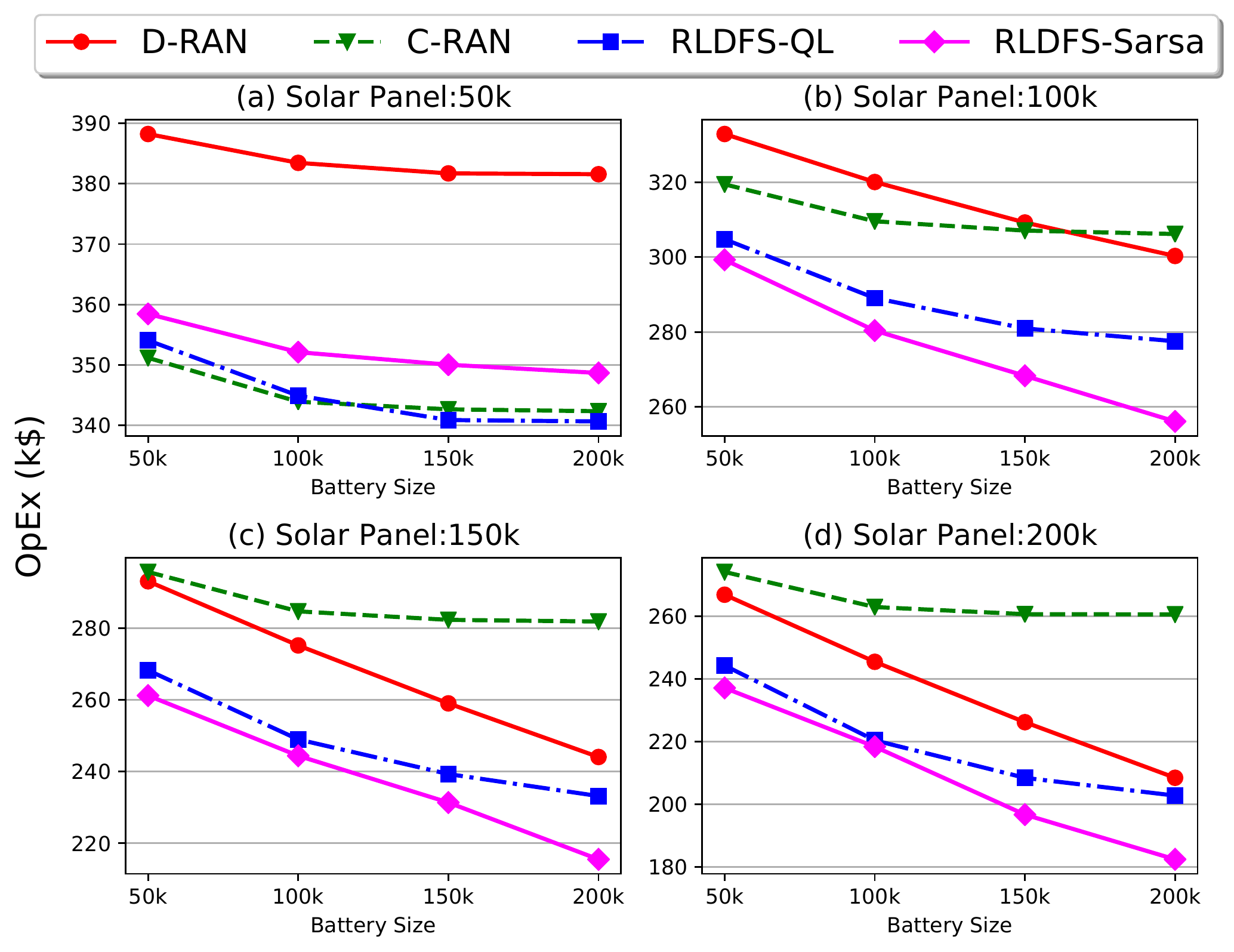}
\caption{\label{fig:panelandbattery} Comparison of methods for different sizes of solar panels and batteries.}
\end{figure}
\par The pvWatts application calculates generated green energy ($G^{y}_{rt}$) from a solar panel \cite{NationalRenewableEnergyLaboratory}. We use the solar radiation data of four different cities (Stockholm, Istanbul, Cairo, Jakarta) with a distinct distribution in a year period. Thus, we can investigate the effect of seasonal change in our model. The other energy consumption parameters are given in Table~\ref{tab:Parameters}. The electricity price values are from the Republic of Turkey Energy Market Regulatory Authorities (EPDK) variable electricity tariff regulation with different price policies according to the time of the day and calculated based on exchange rates as of September 2020 \cite{EPDK}.

\par Table~\ref{tab:rlparam} details the reinforcement learning parameters. The idea to choose a 48 hours window for cost minimization relies on making the decisions to consider the daily variance of traffic loads, harvested energy and electricity prices. Also, we want to extend the decisions between the consecutive days. We evaluate our RL methods with comparison to two approaches. The first one, called distributed-RAN (D-RAN), processes both URLLC and eMBB packets at DUs. The second one, centralized-RAN (C-RAN), handle URLLC packets at DUs to not violate delay requirements and transfer eMBB packets to the CU for cost-efficiency. Figure~\ref{fig:tscity} shows the performance of RLDFS-QL and RLDFS-Sarsa methods according to different cities (solar radiation distributions) and traffic rates. These methods outperform D-RAN and C-RAN approaches in all four cities and under varying traffic intensities (low, medium, high). Besides, our methods perform better with increasing solar radiation rates.
\par Figure~\ref{fig:daily}a provides more insights to the results in Figure 4 by considering the city of Jakarta and medium traffic load. The proposed RLDFS-Sarsa algorithm can consume a higher amount of renewable energy in the noontime; thus, it's the method that has the lowest unstored energy in Figure~\ref{fig:daily}b. Meanwhile, RLDFS-QL algorithm can shift renewable energy usage according to traffic load and the electricity tariffs. C-RAN fails to efficiently use renewable energy owing to not migrating the eMBB packets to the DUs, in the case of lower URLLC traffic loads on DUs.
We further investigate the impact of RES and battery sizing, which is shown in Figure~\ref{fig:panelandbattery}. As observed, RL-based methods have lower costs than the other techniques and their performance improves with larger solar panels and batteries. Adaptation to a larger amount of renewable energy is the main reason for this outcome.

\section{Conclusion}
\par In this paper, we introduce a novel RAN concept that combines energy-efficiency with virtualization that will be applicable to future O-RAN deployments. We propose a reinforcement learning based technique and solve it with two different approaches (Q-learning and Sarsa) to make dynamic function split decisions among DUs and the CU. We also formulate an OpEx minimization problem. Our results show that RL-based solutions make the best use of renewable energy and are cost-efficient. Finally, we present the findings of the impact of different solar panel sizes and the batteries on this network model, which helps to evaluate the feasibility of using RES for an MNO. As a future work, we plan to investigate the optimal RES and battery sizing for an MNO. 

\bibliography{icc2021}

\end{document}